\PassOptionsToPackage{unicode}{hyperref}
\PassOptionsToPackage{hyphens}{url}
\PassOptionsToPackage{dvipsnames,svgnames,x11names}{xcolor}
\documentclass[
]{article}
\usepackage{amsmath,amssymb}
\usepackage{lmodern}
\usepackage{iftex}
\ifPDFTeX
  \usepackage[T1]{fontenc}
  \usepackage[utf8]{inputenc}
  \usepackage{textcomp} % provide euro and other symbols
\else % if luatex or xetex
  \usepackage{unicode-math}
  \defaultfontfeatures{Scale=MatchLowercase}
  \defaultfontfeatures[\rmfamily]{Ligatures=TeX,Scale=1}
\fi
\IfFileExists{upquote.sty}{\usepackage{upquote}}{}
\IfFileExists{microtype.sty}{% use microtype if available
  \usepackage[]{microtype}
  \UseMicrotypeSet[protrusion]{basicmath} % disable protrusion for tt fonts
}{}
\makeatletter
\@ifundefined{KOMAClassName}{% if non-KOMA class
  \IfFileExists{parskip.sty}{%
    \usepackage{parskip}
  }{% else
    \setlength{\parindent}{0pt}
    \setlength{\parskip}{6pt plus 2pt minus 1pt}}
}{% if KOMA class
  \KOMAoptions{parskip=half}}
\makeatother
\usepackage{xcolor}
\usepackage{graphicx}
\makeatletter
\def\maxwidth{\ifdim\Gin@nat@width>\linewidth\linewidth\else\Gin@nat@width\fi}
\def\maxheight{\ifdim\Gin@nat@height>\textheight\textheight\else\Gin@nat@height\fi}
\makeatother
\setkeys{Gin}{width=\maxwidth,height=\maxheight,keepaspectratio}
\makeatletter
\def\fps@figure{htbp}
\makeatother
\newlength{\cslhangindent}
\newlength{\csllabelwidth}
\newlength{\cslentryspacingunit} % times entry-spacing
\newenvironment{CSLReferences}[2] % #1 hanging-ident, #2 entry spacing
 {% don't indent paragraphs
  \setlength{\parindent}{0pt}
  \ifodd #1
  \let\oldpar\par
  \def\par{\hangindent=\cslhangindent\oldpar}
  \fi
  \setlength{\parskip}{#2\cslentryspacingunit}
 }%
 {}
\usepackage{calc}

\ifLuaTeX
\usepackage[bidi=basic]{babel}
\else
\usepackage[bidi=default]{babel}
\fi
\babelprovide[main,import]{american}
\def\languageshorthands#1{}
\ifLuaTeX
  \usepackage{selnolig}  % disable illegal ligatures
\fi
\IfFileExists{bookmark.sty}{\usepackage{bookmark}}{\usepackage{hyperref}}
\IfFileExists{xurl.sty}{\usepackage{xurl}}{} % add URL line breaks if available
\hypersetup{
  pdftitle={GraphNeT: Graph neural networks for neutrino telescope event
reconstruction},
  pdfauthor={Andreas Søgaard, Rasmus F. Ørsøe, Leon Bozianu, Morten
Holm, Kaare Endrup Iversen, Tim Guggenmos, Martin Ha Minh, Philipp
Eller, Troels C. Petersen},
  pdflang={en-US},
  colorlinks=true,
  linkcolor={Maroon},
  filecolor={Maroon},
  citecolor={Blue},
  urlcolor={Blue},
  pdfcreator={LaTeX via pandoc}}

\title{GraphNeT: Graph neural networks for neutrino telescope event
reconstruction}

\usepackage[affil-it]{authblk}
\usepackage{orcidlink}
\author[1%
  \ensuremath\mathparagraph]{Andreas Søgaard%
    \,\orcidlink{0000-0002-0823-056X}\,%
    }
\author[2%
  ]{Rasmus F. Ørsøe%
    \,\orcidlink{0000-0001-8890-4124}\,%
    }
\author[1%
  ]{Leon Bozianu%
    }
\author[1%
  ]{Morten Holm%
    }
\author[1%
  ]{Kaare Endrup Iversen%
    }
\author[2%
  ]{Tim Guggenmos%
    }
\author[2%
  ]{Martin Ha Minh%
    \,\orcidlink{0000-0001-7776-4875}\,%
    }
\author[2%
  ]{Philipp Eller%
    \,\orcidlink{0000-0001-6354-5209}\,%
    }
\author[1%
  ]{Troels C. Petersen%
    \,\orcidlink{0000-0003-0221-3037}\,%
    }

\affil[1]{Niels Bohr Institute, University of Copenhagen, Denmark}
\affil[2]{Technical University of Munich, Germany}
\affil[$\mathparagraph$]{Corresponding author}
\date{16 September 2022}

\begin{document}
\maketitle

\hypertarget{summary}{%
\section{Summary}\label{summary}}

Neutrino telescopes, such as ANTARES
(\protect\hyperlink{ref-ANTARES:2011hfw}{ANTARES Collaboration, 2011b}),
IceCube (\protect\hyperlink{ref-DeepCore}{IceCube Collaboration, 2012},
\protect\hyperlink{ref-Aartsen:2016nxy}{2017}), KM3NeT
(\protect\hyperlink{ref-KM3Net:2016zxf}{KM3NeT Collaboration, 2016}),
and Baikal-GVD (\protect\hyperlink{ref-Baikal-GVD:2018isr}{Baikal-GVD
Collaboration, 2018}) has the science goal of detecting neutrinos and
measuring their properties and origins. Reconstruction at these
experiments is concerned with classifying the type of event or
estimating properties of the interaction.

\texttt{GraphNeT} (\protect\hyperlink{ref-graphnet_zenodo:2022}{Søgaard
et al., 2022}) is an open-source python framework aimed at providing
high quality, user friendly, end-to-end functionality to perform
reconstruction tasks at neutrino telescopes using graph neural networks
(GNNs). \texttt{GraphNeT} makes it fast and easy to train complex models
that can provide event reconstruction with state-of-the-art performance,
for arbitrary detector configurations, with inference times that are
orders of magnitude faster than traditional reconstruction techniques
(\protect\hyperlink{ref-gnn_icecube}{IceCube Collaboration, 2022b}).

GNNs from \texttt{GraphNeT} are flexible enough to be applied to data
from all neutrino telescopes, including future projects such as IceCube
extensions (\protect\hyperlink{ref-IceCube:2016xxt}{IceCube-Gen2
Collaboration, 2017},
\protect\hyperlink{ref-IceCube-Gen2:2020qha}{2021};
\protect\hyperlink{ref-IceCube-PINGU:2014okk}{IceCube-PINGU
Collaboration, 2014}) or P-ONE
(\protect\hyperlink{ref-P-ONE:2020ljt}{P-ONE Collaboration, 2020}). This
means that GNN-based reconstruction can be used to provide
state-of-the-art performance on most reconstruction tasks in neutrino
telescopes, at real-time event rates, across experiments and physics
analyses, with vast potential impact for neutrino and astro-particle
physics.

\hypertarget{statement-of-need}{%
\section{Statement of need}\label{statement-of-need}}

Neutrino telescopes typically consist of thousands of optical modules
(OMs) to detect the Cherenkov light produced from particle interactions
in the detector medium. The number of photo-electrons recorded by the
OMs in each event roughly scales with the energy of the incident
particle, from a few photo-electrons and up to tens of thousands.

Reconstructing the particle type and parameters from individual
recordings (called events) in these experiments is a challenge due to
irregular detector geometry, inhomogeneous detector medium, sparsity of
the data, the large variations of the amount of signal between different
events, and the sheer number of events that need to be reconstructed.

Multiple approaches have been employed, including relatively simple
methods (\protect\hyperlink{ref-Aguilar:2011zz}{ANTARES Collaboration,
2011a}; \protect\hyperlink{ref-IceCube:2022kff}{IceCube Collaboration,
2022a}) that are robust but limited in precision and likelihood-based
methods (\protect\hyperlink{ref-Aartsen:2013bfa}{Aartsen \& others,
2014}; \protect\hyperlink{ref-Abbasi_2013}{Abbasi et al., 2013};
\protect\hyperlink{ref-Ahrens:2003fg}{AMANDA Collaboration, 2004};
\protect\hyperlink{ref-ANTARES:2017ivh}{ANTARES Collaboration, 2017};
\protect\hyperlink{ref-Chirkin:2013avz}{Chirkin, 2013};
\protect\hyperlink{ref-IceCube:2022kff}{IceCube Collaboration, 2022a},
\protect\hyperlink{ref-Aartsen:2013vja}{2014},
\protect\hyperlink{ref-IceCube:2021oqo}{2021b}) that can attain a high
accuracy at the price of high computational cost and detector specific
assumptions.

Recently, machine learning (ML) methods have started to be used, such as
convolutional neural networks (CNNs)
(\protect\hyperlink{ref-Abbasi:2021ryj}{IceCube Collaboration, 2021a};
\protect\hyperlink{ref-Aiello:2020orq}{KM3NeT Collaboration, 2020}) that
are comparably fast, but require detector data being transformed into a
regular pixel or voxel grid. Other approaches get around the geometric
limitations, but increase the computational cost to a similar level as
the traditional likelihood methods
(\protect\hyperlink{ref-Eller:2022xvi}{Eller et al., 2022}).

Instead, GNNs can be thought of as generalised CNNs that work on data
with any geometry, making this paradigm a natural fit for neutrino
telescope data.

The \texttt{GraphNeT} framework provides the end-to-end tools to train
and deploy GNN reconstruction models. \texttt{GraphNeT} leverages
industry-standard tools such as \texttt{pytorch}
(\protect\hyperlink{ref-NEURIPS2019_9015}{Paszke et al., 2019}),
\texttt{PyG}
(\protect\hyperlink{ref-Fey_Fast_Graph_Representation_2019}{Fey \&
Lenssen, 2019}), \texttt{lightning}
(\protect\hyperlink{ref-Falcon_PyTorch_Lightning_2019}{Falcon \& The
PyTorch Lightning team, 2019}), and \texttt{wand}
(\protect\hyperlink{ref-wandb}{Biewald, 2020}) for building and training
GNNs as well as particle physics standard tools such as \texttt{awkward}
(\protect\hyperlink{ref-jim_pivarski_2020_3952674}{Pivarski et al.,
2020}) for handling the variable-size data representing particle
interaction events in neutrino telescopes. The inference speed on a
single GPU allows for processing the full online datastream of current
neutrino telescopes in real-time.

\hypertarget{impact-on-physics}{%
\section{Impact on physics}\label{impact-on-physics}}

\texttt{GraphNeT} provides a common framework for ML developers and
physicists that wish to use the state-of-the-art GNN tools in their
research. By uniting both user groups, \texttt{GraphNeT} aims to
increase the longevity and usability of individual code contributions
from ML developers by building a general, reusable software package
based on software engineering best practices, and lowers the technical
threshold for physicists that wish to use the most performant tools for
their scientific problems.

The \texttt{GraphNeT} models can improve event classification and yield
very accurate reconstruction, e.g., for low energy neutrinos observed in
IceCube. Here, a GNN implemented in \texttt{GraphNeT} was applied to the
problem of neutrino oscillations in IceCube, leading to significant
improvements in both energy and angular reconstruction in the energy
range relevant to oscillation studies
(\protect\hyperlink{ref-gnn_icecube}{IceCube Collaboration, 2022b}).
Furthermore, it was shown that the GNN could improve muon vs.~neutrino
classification and thereby the efficiency and purity of a neutrino
sample for such an analysis.

Similarly, improved angular reconstruction has a great impact on, e.g.,
neutrino point source analyses.

Finally, the fast (order millisecond) reconstruction allows for a whole
new type of cosmic alerts at lower energies, which were previously
unfeasible. GNN-based reconstruction makes it possible to identify low
energy (\textless{} 10 TeV) neutrinos and monitor their rate, direction,
and energy in real-time. This will enable cosmic neutrino alerts based
on such neutrinos for the first time ever, despite a large background of
neutrinos that are not of cosmic origin.

\hypertarget{usage}{%
\section{Usage}\label{usage}}

\begin{figure}
\centering
\includegraphics{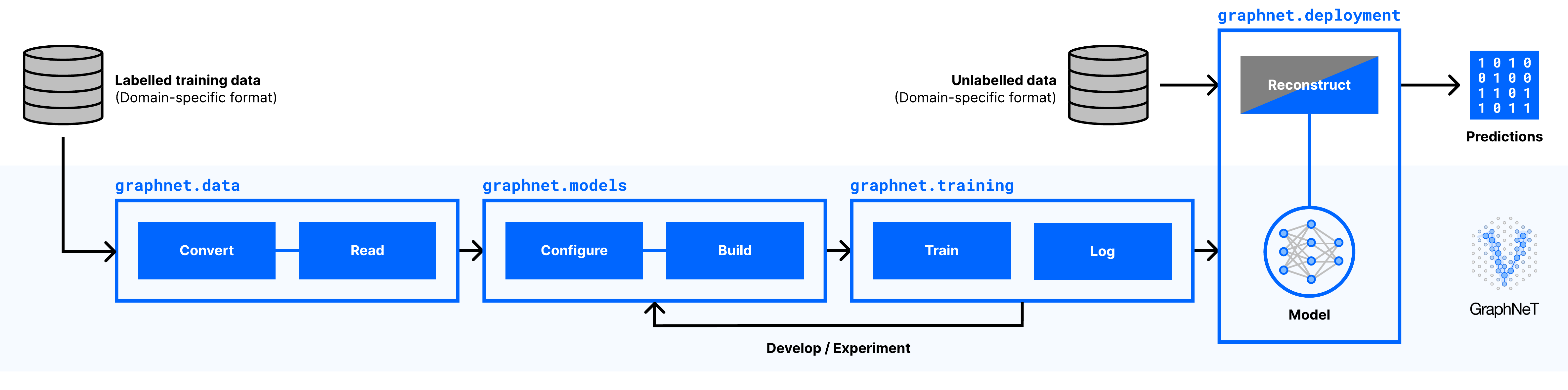}
\caption{High-level overview of a typical workflow using
\texttt{GraphNeT}: \texttt{graphnet.data} enables converting
domain-specific data to industry-standard, intermediate file formats and
reading this data; \texttt{graphnet.models} allows for configuring and
building complex GNN models using simple, physics-oriented components;
\texttt{graphnet.training} manages model training and experiment
logging; and finally, \texttt{graphnet.deployment} allows for using
trained models for inference in domain-specific reconstruction
chains.\label{fig:flowchart}}
\end{figure}

\texttt{GraphNeT} comprises a number of modules providing the necessary
tools to build workflow from ingesting raw training data in
domain-specific formats to deploying trained models in domain-specific
reconstruction chains, as illustrated in \autoref{fig:flowchart}.

\texttt{graphnet.models} provides modular components subclassing
\texttt{torch.nn.Module}, meaning that users only need to import a few,
existing, purpose-built components and chain them together to form a
complete GNN. ML developers can contribute to \texttt{GraphNeT} by
extending this suite of model components --- through new layer types,
physics tasks, graph connectivities, etc. --- and experiment with
optimising these for different reconstruction tasks using experiment
tracking.

These models are trained using \texttt{graphnet.training} on data
prepared using \texttt{graphnet.data}, to satisfy the high I/O loads
required when training ML models on large batches of events, which
domain-specific neutrino physics data formats typically do not allow.

Trained models are deployed to a domain-specific reconstruction chain,
yielding predictions, using the components in
\texttt{graphnet.deployment}. This can either be through model files or
container images, making deployment as portable and dependency-free as
possible.

By splitting up the GNN development as in \autoref{fig:flowchart},
\texttt{GraphNeT} allows physics users to interface only with high-level
building blocks or pre-trained models that can be used directly in their
reconstruction chains, while allowing ML developers to continuously
improve and expand the framework's capabilities.

\hypertarget{acknowledgements}{%
\section{Acknowledgements}\label{acknowledgements}}

Andreas Søgaard has received funding from the European Union's Horizon
2020 research and innovation programme under the Marie Skłodowska-Curie
grant agreement No.~890778. The work of Rasmus Ørsøe was partly
performed in the framework of the PUNCH4NFDI consortium supported by DFG
fund ``NFDI 39/1'', Germany.

\hypertarget{references}{%
\section*{References}\label{references}}
\addcontentsline{toc}{section}{References}

\hypertarget{refs}{}
\begin{CSLReferences}{1}{0}
\leavevmode\vadjust pre{\hypertarget{ref-Aartsen:2013bfa}{}}%
Aartsen, M. G., \& others. (2014). {Improvement in Fast Particle Track
Reconstruction with Robust Statistics}. \emph{Nucl. Instrum. Meth. A},
\emph{736}, 143--149. \url{https://doi.org/10.1016/j.nima.2013.10.074}

\leavevmode\vadjust pre{\hypertarget{ref-Abbasi_2013}{}}%
Abbasi, R., Abdou, Y., Ackermann, M., Adams, J., Aguilar, J. A., Ahlers,
M., Altmann, D., Andeen, K., Auffenberg, J., Bai, X., \& al., et.
(2013). An improved method for measuring muon energy using the truncated
mean of dE/dx. \emph{Nucl. Instum. Meth. A}, \emph{703}, 190--198.

\leavevmode\vadjust pre{\hypertarget{ref-Ahrens:2003fg}{}}%
AMANDA Collaboration. (2004). {Muon track reconstruction and data
selection techniques in AMANDA}. \emph{Nucl. Instrum. Meth. A},
\emph{524}, 169--194. \url{https://doi.org/10.1016/j.nima.2004.01.065}

\leavevmode\vadjust pre{\hypertarget{ref-Aguilar:2011zz}{}}%
ANTARES Collaboration. (2011a). {A fast algorithm for muon track
reconstruction and its application to the ANTARES neutrino telescope}.
\emph{Astropart. Phys.}, \emph{34}, 652--662.
\url{https://doi.org/10.1016/j.astropartphys.2011.01.003}

\leavevmode\vadjust pre{\hypertarget{ref-ANTARES:2011hfw}{}}%
ANTARES Collaboration. (2011b). {ANTARES: the first undersea neutrino
telescope}. \emph{Nucl. Instrum. Meth. A}, \emph{656}, 11--38.
\url{https://doi.org/10.1016/j.nima.2011.06.103}

\leavevmode\vadjust pre{\hypertarget{ref-ANTARES:2017ivh}{}}%
ANTARES Collaboration. (2017). {An algorithm for the reconstruction of
neutrino-induced showers in the ANTARES neutrino telescope}.
\emph{Astron. J.}, \emph{154}(6), 275.
\url{https://doi.org/10.3847/1538-3881/aa9709}

\leavevmode\vadjust pre{\hypertarget{ref-Baikal-GVD:2018isr}{}}%
Baikal-GVD Collaboration. (2018). {Baikal-GVD: status and prospects}.
\emph{EPJ Web Conf.}, \emph{191}, 01006.
\url{https://doi.org/10.1051/epjconf/201819101006}

\leavevmode\vadjust pre{\hypertarget{ref-wandb}{}}%
Biewald, L. (2020). \emph{Experiment tracking with weights and biases}.
\url{https://www.wandb.com/}

\leavevmode\vadjust pre{\hypertarget{ref-Chirkin:2013avz}{}}%
Chirkin, D. (2013). {Event reconstruction in IceCube based on direct
event re-simulation}. \emph{{33rd International Cosmic Ray Conference}}.

\leavevmode\vadjust pre{\hypertarget{ref-Eller:2022xvi}{}}%
Eller, P., Fienberg, A., Weldert, J., Wendel, G., Böser, S., \& Cowen,
D. F. (2022). \emph{{A flexible event reconstruction based on machine
learning and likelihood principles}}.
\url{https://arxiv.org/abs/2208.10166}

\leavevmode\vadjust pre{\hypertarget{ref-Falcon_PyTorch_Lightning_2019}{}}%
Falcon, W., \& The PyTorch Lightning team. (2019). \emph{{PyTorch
Lightning}} (Version 1.4) {[}Computer software{]}.
\url{https://doi.org/10.5281/zenodo.3828935}

\leavevmode\vadjust pre{\hypertarget{ref-Fey_Fast_Graph_Representation_2019}{}}%
Fey, M., \& Lenssen, J. E. (2019). \emph{{Fast Graph Representation
Learning with PyTorch Geometric}}.
\url{https://github.com/pyg-team/pytorch_geometric}

\leavevmode\vadjust pre{\hypertarget{ref-IceCube:2022kff}{}}%
IceCube Collaboration. (2022a). \emph{{Low Energy Event Reconstruction
in IceCube DeepCore}}. \url{https://arxiv.org/abs/2203.02303}

\leavevmode\vadjust pre{\hypertarget{ref-DeepCore}{}}%
IceCube Collaboration. (2012). {The design and performance of IceCube
DeepCore}. \emph{Astroparticle Physics}, \emph{35}(10), 615--624.

\leavevmode\vadjust pre{\hypertarget{ref-Aartsen:2013vja}{}}%
IceCube Collaboration. (2014). {Energy Reconstruction Methods in the
IceCube Neutrino Telescope}. \emph{JINST}, \emph{9}, P03009.
\url{https://doi.org/10.1088/1748-0221/9/03/P03009}

\leavevmode\vadjust pre{\hypertarget{ref-Aartsen:2016nxy}{}}%
IceCube Collaboration. (2017). {The IceCube Neutrino Observatory:
Instrumentation and Online Systems}. \emph{JINST}, \emph{12}(03),
P03012. \url{https://doi.org/10.1088/1748-0221/12/03/P03012}

\leavevmode\vadjust pre{\hypertarget{ref-Abbasi:2021ryj}{}}%
IceCube Collaboration. (2021a). {A Convolutional Neural Network based
Cascade Reconstruction for the IceCube Neutrino Observatory}.
\emph{JINST}, \emph{16}, P07041.
\url{https://doi.org/10.1088/1748-0221/16/07/P07041}

\leavevmode\vadjust pre{\hypertarget{ref-IceCube:2021oqo}{}}%
IceCube Collaboration. (2021b). {A muon-track reconstruction exploiting
stochastic losses for large-scale Cherenkov detectors}. \emph{JINST},
\emph{16}(08), P08034.
\url{https://doi.org/10.1088/1748-0221/16/08/P08034}

\leavevmode\vadjust pre{\hypertarget{ref-gnn_icecube}{}}%
IceCube Collaboration. (2022b). \emph{{Graph Neural Networks for
Low-Energy Event Classification \& Reconstruction in IceCube}}.
\url{https://arxiv.org/abs/2209.03042}

\leavevmode\vadjust pre{\hypertarget{ref-IceCube:2016xxt}{}}%
IceCube-Gen2 Collaboration. (2017). {PINGU: A Vision for Neutrino and
Particle Physics at the South Pole}. \emph{J. Phys. G}, \emph{44}(5),
054006. \url{https://doi.org/10.1088/1361-6471/44/5/054006}

\leavevmode\vadjust pre{\hypertarget{ref-IceCube-Gen2:2020qha}{}}%
IceCube-Gen2 Collaboration. (2021). {IceCube-Gen2: the window to the
extreme Universe}. \emph{J. Phys. G}, \emph{48}(6), 060501.
\url{https://doi.org/10.1088/1361-6471/abbd48}

\leavevmode\vadjust pre{\hypertarget{ref-IceCube-PINGU:2014okk}{}}%
IceCube-PINGU Collaboration. (2014). \emph{{Letter of Intent: The
Precision IceCube Next Generation Upgrade (PINGU)}}.
\url{https://arxiv.org/abs/1401.2046}

\leavevmode\vadjust pre{\hypertarget{ref-KM3Net:2016zxf}{}}%
KM3NeT Collaboration. (2016). {Letter of intent for KM3NeT 2.0}.
\emph{J. Phys. G}, \emph{43}(8), 084001.
\url{https://doi.org/10.1088/0954-3899/43/8/084001}

\leavevmode\vadjust pre{\hypertarget{ref-Aiello:2020orq}{}}%
KM3NeT Collaboration. (2020). {Event reconstruction for KM3NeT/ORCA
using convolutional neural networks}. \emph{JINST}, \emph{15}(10),
P10005. \url{https://doi.org/10.1088/1748-0221/15/10/P10005}

\leavevmode\vadjust pre{\hypertarget{ref-NEURIPS2019_9015}{}}%
Paszke, A., Gross, S., Massa, F., Lerer, A., Bradbury, J., Chanan, G.,
Killeen, T., Lin, Z., Gimelshein, N., Antiga, L., Desmaison, A., Kopf,
A., Yang, E., DeVito, Z., Raison, M., Tejani, A., Chilamkurthy, S.,
Steiner, B., Fang, L., \ldots{} Chintala, S. (2019). PyTorch: An
imperative style, high-performance deep learning library. In H. Wallach,
H. Larochelle, A. Beygelzimer, F. dAlché-Buc, E. Fox, \& R. Garnett
(Eds.), \emph{Advances in neural information processing systems 32} (pp.
8024--8035). Curran Associates, Inc.
\url{http://papers.neurips.cc/paper/9015-pytorch-an-imperative-style-high-performance-deep-learning-library.pdf}

\leavevmode\vadjust pre{\hypertarget{ref-jim_pivarski_2020_3952674}{}}%
Pivarski, J., Escott, C., Smith, N., Hedges, M., Proffitt, M., Escott,
C., Nandi, J., Rembser, J., bfis, benkrikler, Gray, L., Davis, D.,
Schreiner, H., Nollde, Fackeldey, P., \& Das, P. (2020).
\emph{Scikit-hep/awkward-array: 0.13.0} (Version 0.13.0) {[}Computer
software{]}. Zenodo. \url{https://doi.org/10.5281/zenodo.3952674}

\leavevmode\vadjust pre{\hypertarget{ref-P-ONE:2020ljt}{}}%
P-ONE Collaboration. (2020). {The Pacific Ocean Neutrino Experiment}.
\emph{Nature Astron.}, \emph{4}(10), 913--915.
\url{https://doi.org/10.1038/s41550-020-1182-4}

\leavevmode\vadjust pre{\hypertarget{ref-graphnet_zenodo:2022}{}}%
Søgaard, A., Ørsøe, R. F., Bozianu, L., Holm, M., Iversen, K. E.,
Guggenmos, T., Minh, M. H., \& Eller, P. (2022). \emph{GraphNeT}.
Zenodo. \url{https://doi.org/10.5281/zenodo.6720188}

\end{CSLReferences}

\end{document}